\documentclass[twocolumn,showpacs,preprintnumbers,amsmath,amssymb,superscriptaddress]{revtex4}
\usepackage{graphicx}
\usepackage{dcolumn}
\usepackage{bm}
\usepackage{natbib}

\def\ltsima{$\; \buildrel < \over \sim \;$}
\def\simlt{\lower.5ex\hbox{\ltsima}}
\def\gtsima{$\; \buildrel > \over \sim \;$}
\def\simgt{\lower.5ex\hbox{\gtsima}}
\begin{document}

\title{The redshift sensitivities of dark energy surveys}

\author{Fergus Simpson}
\email{frgs@ast.cam.ac.uk} \affiliation{Institute of Astronomy,
University of Cambridge, Madingley Road, Cambridge CB3 0HA}
\author{Sarah Bridle}
 \affiliation{Department of Physics and Astronomy, University College London, Gower Street,
London WC1E 6BT}
\date{\today}
\newcommand{\ud}{\mathrm{d}}
\begin{abstract}
Great uncertainty surrounds dark energy, both in terms of its
physics, and the choice of methods by which the problem should be
addressed. Here we quantify the redshift sensitivities offered by
different techniques. We focus on the three methods most adept at
constraining $w$, namely supernovae, cosmic shear, and baryon
oscillations. For each we provide insight into the family of
$w(z)$ models which are permitted for a particular constraint on
either $w=w_0$ or $w=w_0+w_a(1-a)$. Our results are in the form of
``weight functions", which describe the fitted model parameters as
a weighted average over the true functional form. For example, we
find the recent best-fit from the Supernovae Legacy Survey
($w=-1.023$) corresponds to the average value of $w(z)$ over the
range $0<z<0.4$. Whilst there is a strong dependence on the choice
of priors, each cosmological probe displays distinctive
characteristics in their redshift sensitivities. In the case of
proposed future surveys, a SNAP-like supernova survey probes a
mean redshift of $z \sim 0.3$, with baryon oscillations and cosmic
shear at $z \sim 0.6$. If we consider the evolution of $w$,
sensitivities shift to slightly higher redshift. Finally, we find
that the weight functions may be expressed as a weighted average
of the popular ``principal components".
\end{abstract}
\maketitle
\section{\label{sec:Intro}Introduction}
Following the rapid accumulation of experimental evidence
 (including
\cite{1998AJ....116.1009R,1999ApJ...517..565P,
2004ApJ...607..665R,hoekstra-2005-,2005MNRAS.362..505C,
2001PhRvD..63d2001L, 2003ApJS..148..175S, 2004MNRAS.353..457A}),
it has become widely accepted that the Universe is experiencing a
period of accelerated expansion. In the context of general
relativity, this implies that the current cosmological dynamics
are dominated by a component with negative pressure. There is
currently no satisfactory theoretical explanation despite the
proliferation of models that have been suggested. Theories range
from vacuum energy, through scalar fields, to modifications of
general relativity, although none have yet produced satisfactory
solutions to the theoretical issues. By revealing the behavior of
this dark energy, not only might we foresee the ultimate fate of
the Universe, but this may be the first step into a new field of
physics.

Dark energy is often parameterised by its equation of state, the
pressure to density ratio $w=p/\rho$. This is believed to
summarise the main effect of dark energy on the observable
universe. This equation of state is expected to change with time,
and therefore redshift (with the notable exception of the
cosmological constant).

\begin{figure} [b]
\includegraphics[width=3.5in]{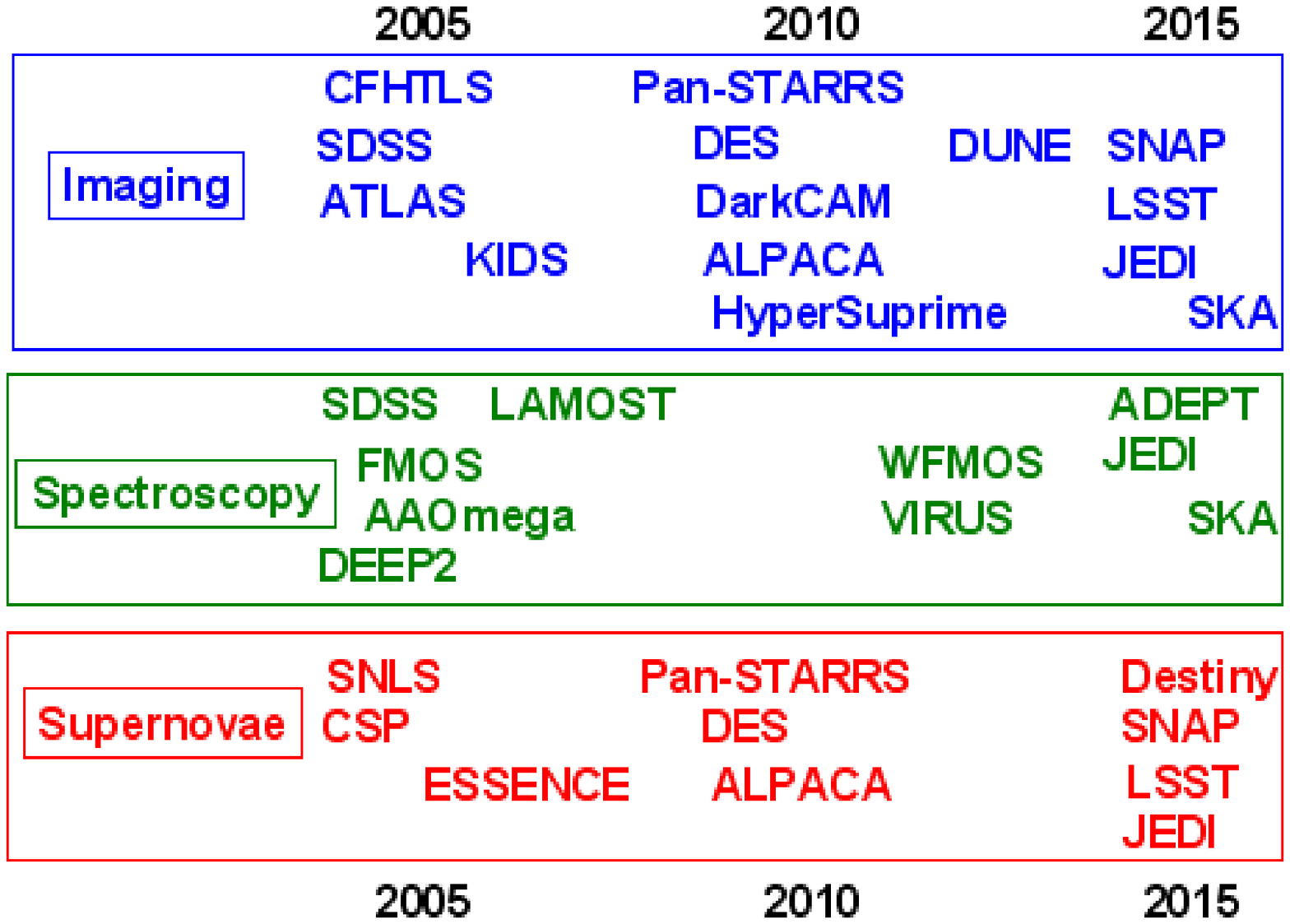}
\caption{\label{fig:expts} Sketch illustrating the large number of
experiments proposing to measure dark energy using the probes
discussed in this paper.}
\end{figure}

Since dark energy is generally considered to be the biggest
problem facing cosmology today, there are a multitude of proposed
surveys which aim to find out more about its nature (summarised in
Fig.~\ref{fig:expts}). Decisions need to be made regarding which
surveys to carry out, given limited funds. Since the proposed
surveys reduce the random uncertainties by more than an order of
magnitude, we have to admit the possibility that systematic
uncertainties may come to dominate the constraints on dark energy,
for some cosmological probes. Unfortunately it is extremely
difficult to quantify these systematic uncertainties, and a
rigorous treatment of these lie beyond the scope of this paper. We
focus on a comparison of three different probes (supernovae,
cosmic shear and baryon oscillations), putting them on an equal
footing and assessing their theoretical complementarity.

Supernova light curves offer the most mature probe of dark energy,
with a number of surveys currently in progress. They are used to
infer the expansion history through measurements of the luminosity
distance, and its variation with redshift.

The gravitational deflection of light slightly warps the images of
distant galaxies. These distortions are sensitive to dark energy
via the distance-redshift relation, and to a lesser extent by the
growth of structure. Current constraints are modest, with an upper
bound of $w \lesssim -0.5$ \cite{hoekstra-2005-} (roughly
$2\sigma$).

Oscillatory features in the galaxy power spectrum have recently
been seen by two redshift surveys
\cite{2005ApJ...633..560E,2005MNRAS.362..505C}. By resolving these
with higher precision and over a range of redshifts, competitive
constraints on $w$ could be achieved
\cite{2003ApJ...594..665B,2003ApJ...598..720S,2004NewAR..48.1063B}.

The potential of a particular cosmological probe can be assessed
in a variety of ways. The most common approach has been to assume
that the dark energy equation of state takes a simple form as a
function of redshift, and then the predicted constraints on the
parameters of the function are calculated. It is highly
challenging for even the ambitious proposed experiments to
constrain the detailed evolution of the equation of state. Each
probe can typically only give one or two parameters
\cite{2005PhRvD..72d3509L,2004MNRAS.348..603S}, and therefore the
equation of state is often expanded to zeroth or first order in
redshift. (Such that $w(z)=w_0$ or $w(z) = w_0 + w_1 z$ or into
the more theoretically motivated form $w(z) = w_0 + w_a(1-a)$ in
which the dark energy has a value $w_0$ today, and gradually
approaches $w_0+w_a$ toward higher redshift). These constraints
are important for comparing the potential of different surveys and
probes.

However, it has been realised that the resulting contour plots do
not tell the whole story of how sensitive each probe is as a
function of redshift, or how many independent pieces of
information each will yield on dark energy. The principal
component approach assesses both of these points
\cite{2003PhRvL..90c1301H,2005ASPC..339..107K,crittenden-2005-}.
This is achieved by attempting to measure the equation of state in
narrow bins in redshift, and diagonalising the resulting
correlated error matrix.

The weight function approach
\cite{2003MNRAS.343..533D,2005PhRvD..71h3501S} aims to address a
slightly different question. It allows us to interpret fitted
parameters, such as $w_0$, in terms of a weighted redshift average
over the true, potentially varying, equation of state. The
resulting weight function also gives an impression of the
sensitivity of each probe as a function of redshift.

Unfortunately, conclusions on all of the above issues depend on
the assumptions made about dark energy and other priors. Priors
have to be placed on the other cosmological parameters as well as
on the range of possibilities allowed for the dark energy.

Here we attempt to bring all of these issues together and compare
the three cosmological probes listed above on an equal basis. We
consider the ability of the various probes to disentangle multiple
pieces of information on the dark energy using a PCA analysis; and
not only do our weight functions look similar to the principal
components, but they are simply related. The redshift
sensitivities of different probes are compared, in the case of
both constant and evolving equation of state parameterisations.
The effect of different priors on cosmological parameters is
explored. Finally, we assess the bias which can arise when fitting
other cosmological parameters.

\section{Concept}

Intuitively one might imagine that slight deviations in $w$ would
be reflected by a \emph{representative} shift of our observed
value. However, as we shall see, this is not necessarily the case.
A particularly extreme example was highlighted by Maor et al.
\cite{2002PhRvD..65l3003M}, in which a quintessence model of
$w(z)\geq-0.7$ was shown to provide a best-fit of $w<-1$. This
fundamentally arises from (a) the incorrect parameterisation of
$w$ and (b) our incomplete knowledge of other cosmological
parameters. By improving our data, these effects will be
suppressed but never eliminated.

Inevitably, results will emerge which assume a time-independent
value for $w$, or at best permitting some pre-determined variation
such as $w=w_0+w_a(1-a)$
\cite{chevallier-2001-10,2003PhRvL..90i1301L}. The aim here is to
provide insight into the meaning of these results, within the
context of dark energy actually exhibiting an arbitrary $w(z)$. In
doing so, we quantify the redshift sensitivities of different
surveys. Central to our approach will be expressing the best-fit
value of $w$ as a weighted integral over the true $w(z)$. The key
ingredient is the ``weight function" $\Phi(z)$ defined such that
\begin{equation} \label{eq:phi}
w^{fit}=\int \Phi(z) w(z) \ud z \,\,\,.
\end{equation}

\noindent This weight function can be interpreted as the
redshift-sensitivity of the observation. In principle, $\Phi(z)$
depends on $w(z)$, although we find that in practice the
dependence is weak (see Appendix \ref{sec:w08}). In this paper all
$\Phi(z)$ functions are calculated using the fiducial model
$w(z)=-1$, unless stated otherwise.

To derive the weight functions, we adopt a different path from
previous work. This provides greater flexibility, and allows
marginalisation over other parameters. We begin by permitting $w$
to adopt an independent value within each infinitesimal redshift
bin. The eigenvectors of the resulting Fisher matrix, marginalised
over the relevant parameters, provide the `principal components',
or eigenmodes, of the survey. These orthonormal functions $e_i(z)$
allow us to express $w(z)$ in the form
\begin{equation} \label{eq:pcas}
w(z)=\sum_i \alpha_i e_i(z)
\end{equation}
\noindent such that the errors $\sigma(\alpha_i)$ are
uncorrelated. Further details of this approach are explored by
Huterer \& Starkman \cite{2003PhRvL..90c1301H}. Weight functions
are most readily derived from the principal components (hereafter
PCAs) of a given survey. For example, in the case of fitting a
constant $w$, using chi-squared minimisation we find
\begin{equation}
\label{eq:phipca} \Phi(z)= \frac{\sum_i{e_i(z)\int e_i(z') \ud
z'/\sigma^2(\alpha_i)}}{\sum_j (\int e_j(z'') \ud z'')^2
/\sigma^2(\alpha_j)} \,\,.
\end{equation}

The weight function $\Phi(z)$ is a sum of the eigenmodes weighted
with the strength by which their coefficients are determined, and
so is a representation of the sensitivity. For further details see
Appendix \ref{sec:WtPCA}. Note that in general, any complete
orthonormal set such as these principal components, are their own
weight functions.

\begin{figure*} [t]
\begin{tabular}{ll}
 \includegraphics[width=2.9in]{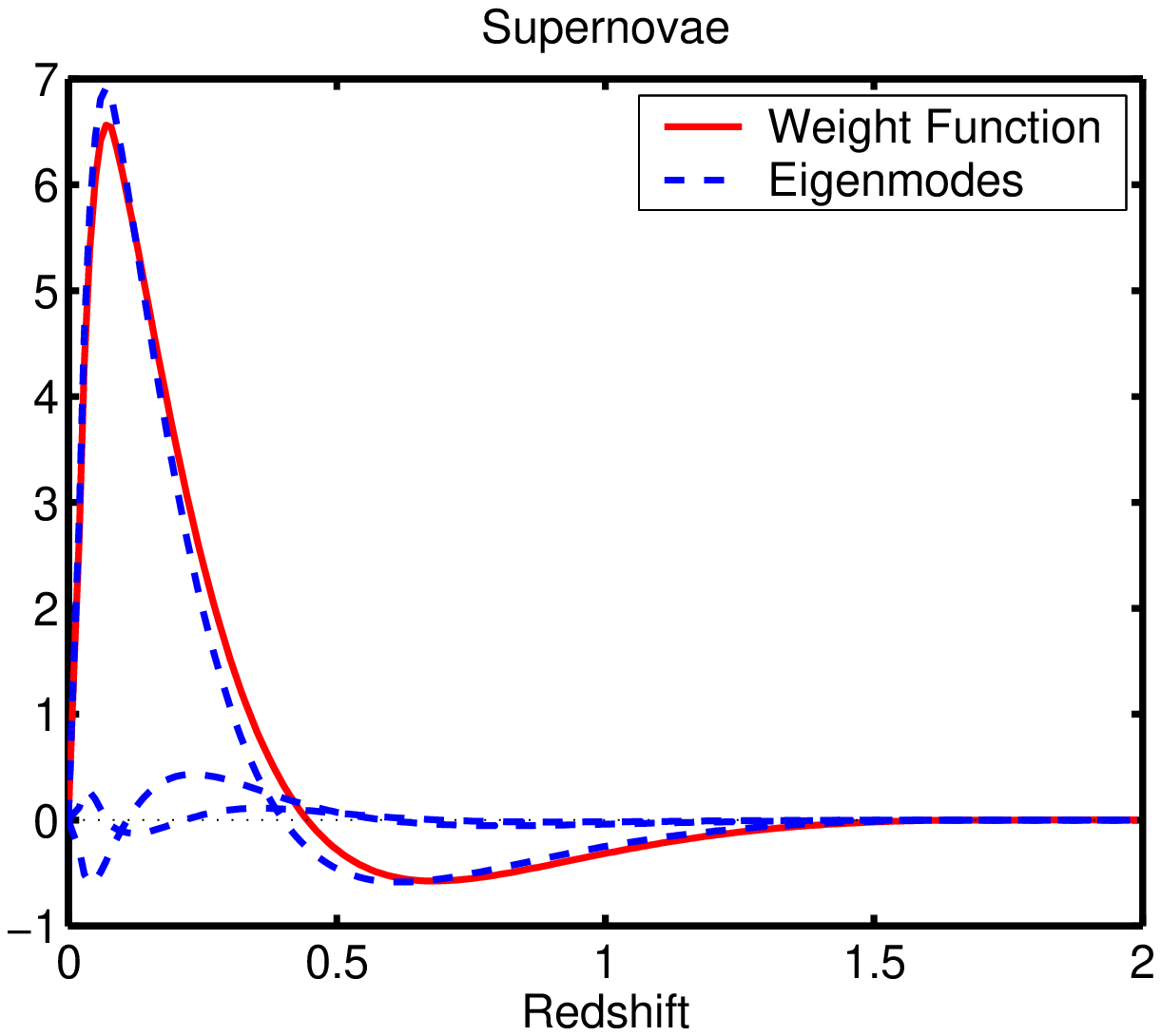}
& \includegraphics[width=3in]{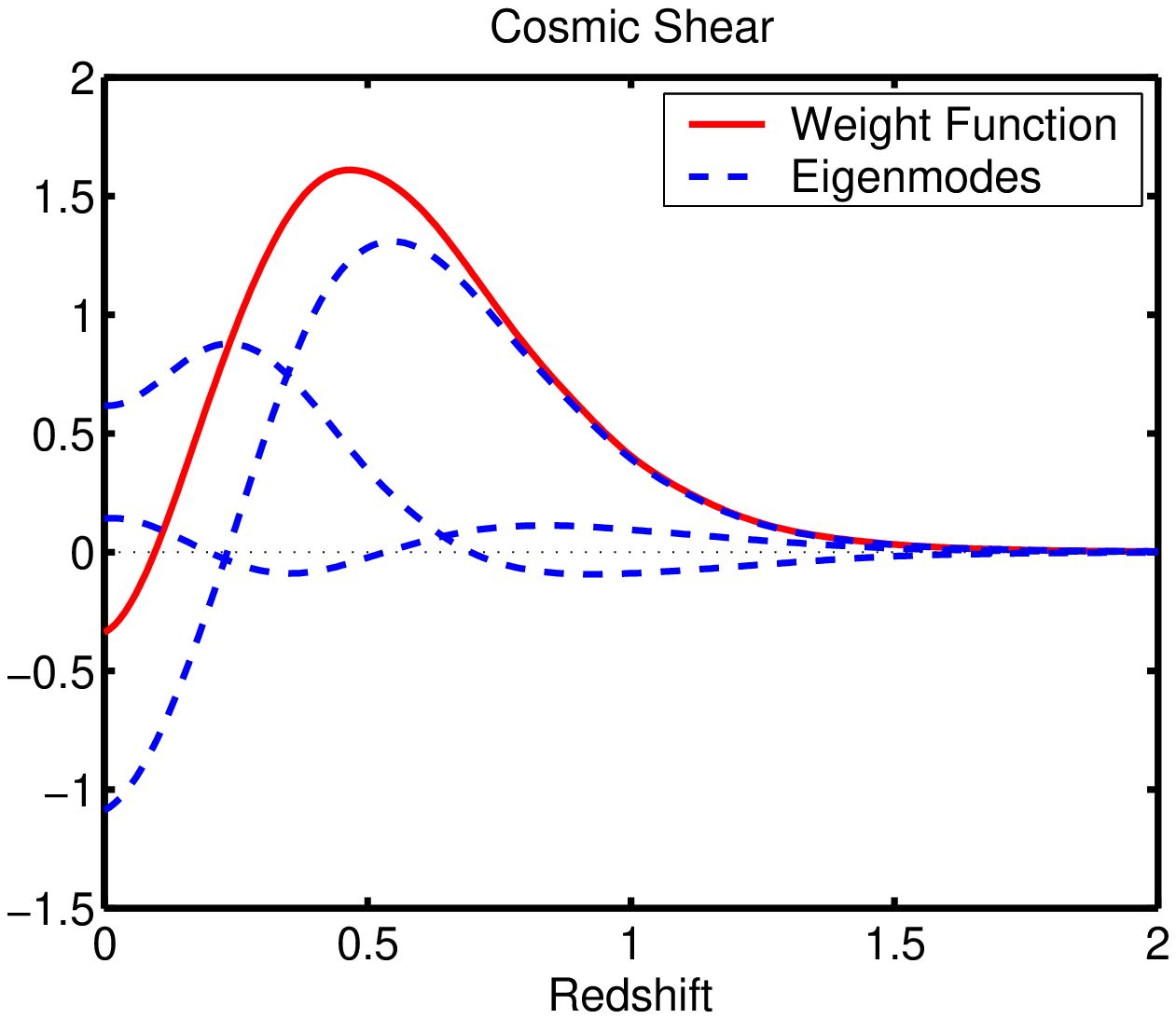}\\
\includegraphics[width=3in]{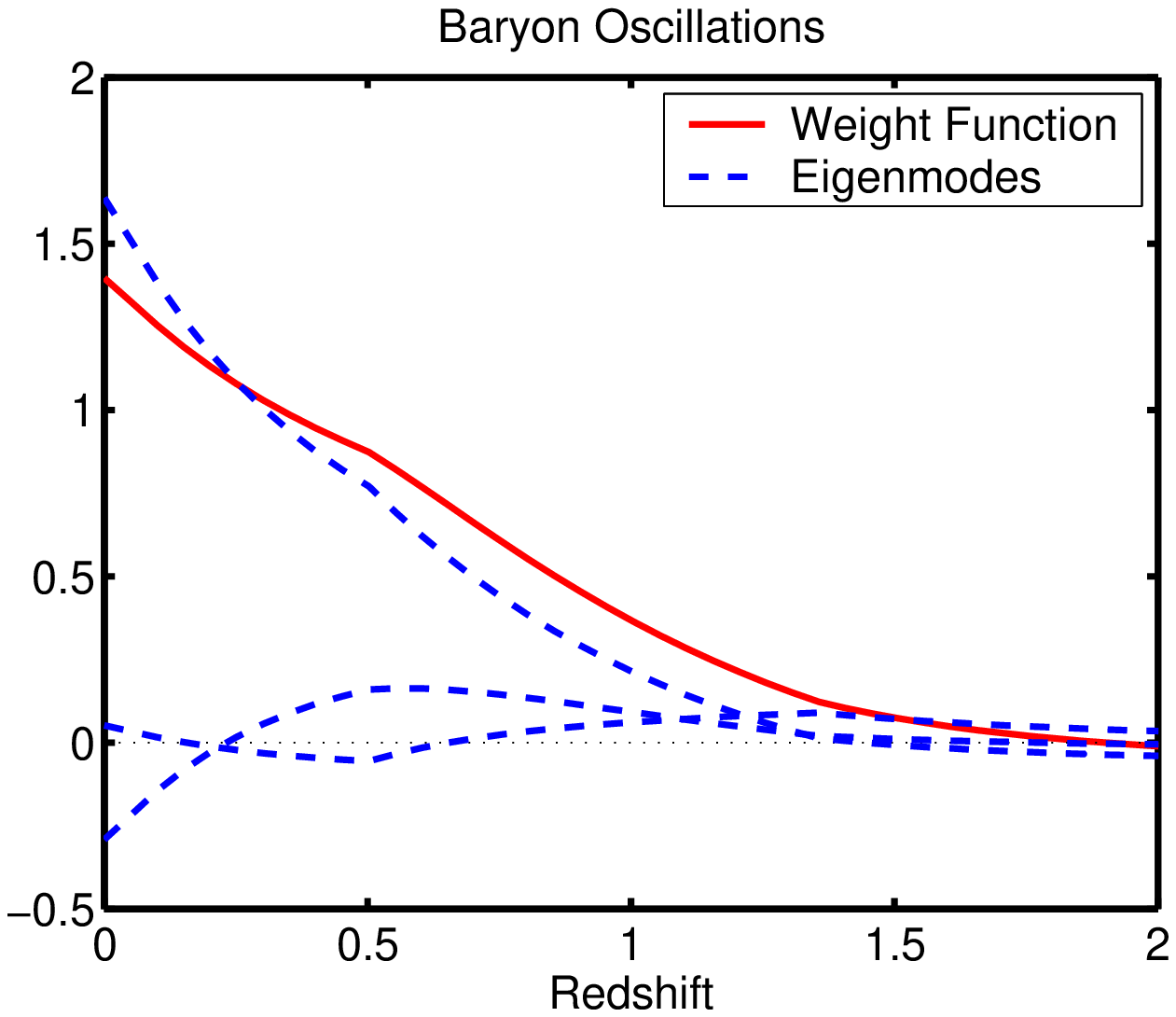}
& \includegraphics[width=3in]{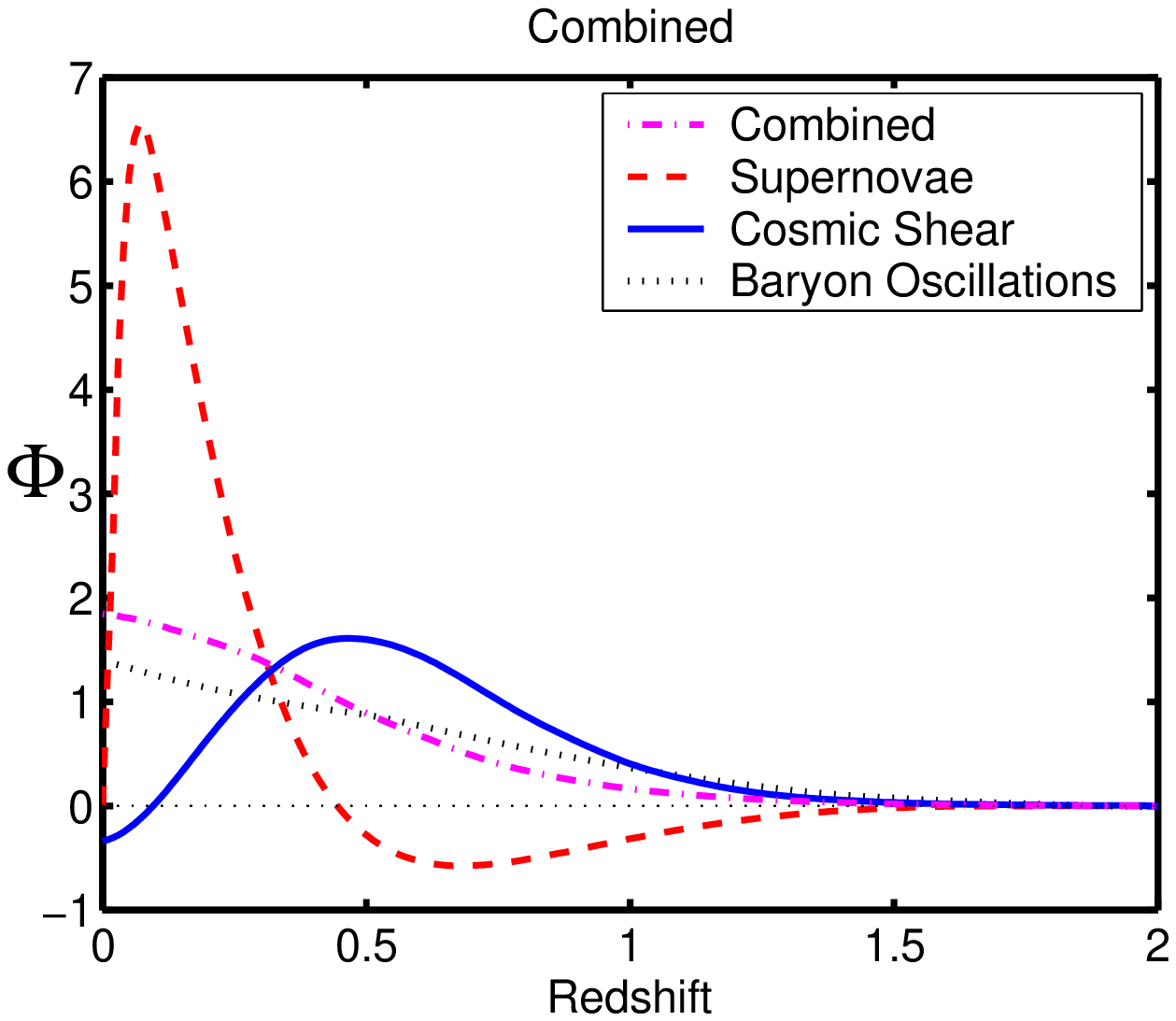}\\
\end{tabular}
\caption{ \emph{Top left:} \label{fig:snaePCA} Supernovae. First
three PCA eigenmodes (dashed lines) weighted in accordance with
equation (\ref{eq:phipca}), the sum of which gives the weight
function (solid line) for a constant $w$ fit. For a SNAP-like
survey. \emph{Top right:} \label{fig:shearPCA} Similarly for
cosmic shear, whose weight function has significant contributions
from the higher modes. For a SNAP-like distribution of source
galaxies (weight functions and PCAs are independent of the survey
area). \emph{Bottom left:} \label{fig:baoPCA} Same for baryon
oscillations. The weight function is distinctive in its
simplicity, with a slight discontinuity in the gradient
corresponding to the edge of the lower redshift bin. Survey
parameters correspond to the proposed WFMOS survey. \emph{Bottom
right:} \label{fig:Const} Here we superpose the weight functions
of the previous plots, allowing for a more detailed comparison.
 In addition, the dot-dash line represents the sensitivity of a combined analysis.}
\end{figure*}

\section{The Weight Functions}
In this section we consider each of the three main cosmological
probes of dark energy, deriving PCAs and weight functions for
fitting a constant dark energy equation of state. We vary the main
relevant cosmological parameters, marginalising over them with
flat priors on each (this improves on our previous work). In
section \ref{sec:Prior} we examine the effect of these priors.

A flat universe is assumed throughout this paper. Whilst the
possibility of non-zero curvature should not be completely
ignored, a flat universe still remains the most likely theoretical
option, and will be used to provide the tightest constraints on
$w$. Curvature is trivially incorporated into this approach, and
the implications for PCAs and weight functions may be explored in
future work.

Each of the following subsections briefly summarises the relevant
physics, outlines the fiducial survey parameters, and discusses
the resulting PCAs and weight functions from
Fig.~\ref{fig:snaePCA}. Note that in Appendix
\ref{sec:scalefactor} this collection of weight functions are
re-expressed in terms of the scale factor rather than redshift.

\subsection{Supernovae}

Our fiducial survey consists of 2000 supernovae uniformly
distributed from $z=0.1-2$, with a further 300 in the range $0.03
< z < 0.08$, as anticipated from the Nearby Supernova Factory. We
calculate the PCAs and weight functions from measurements of the
observed magnitudes, $m$,
\begin{equation}
m - \mathcal{M} = 5 \log_{10} D_L
\end{equation}
\noindent where $\mathcal{M}$ incorporates the intrinsic magnitude
and the Hubble parameter, and the redshift dependence of the
luminosity distance $D_L$ provides our grasp on the equation of
state.

We apply a flat prior to the parameters $\mathcal{M}$ and
$\Omega_{\rm{m}}$. Following Linder \& Huterer
\cite{2003PhRvD..67h1303L}, we include an irreducible uncertainty
of the form $\ud m = 0.02(1+z)/2.7$ in bins of width $\Delta
z=0.1$, to mimic systematics.

The PCA eigenvalues deteriorate quite sharply, and higher modes
are penalised for their oscillatory nature, so the weight function
looks similar to the first PCA eigenmode. The weight function
peaks at around $z \sim 0.1$ and has a smaller negative tail
beyond $z>0.45$. We consider the mean of $|\Phi(z)|$ to be a good
benchmark, and in this case find a value of 0.28.

\begin{figure*}[t]
\includegraphics[width=3in]{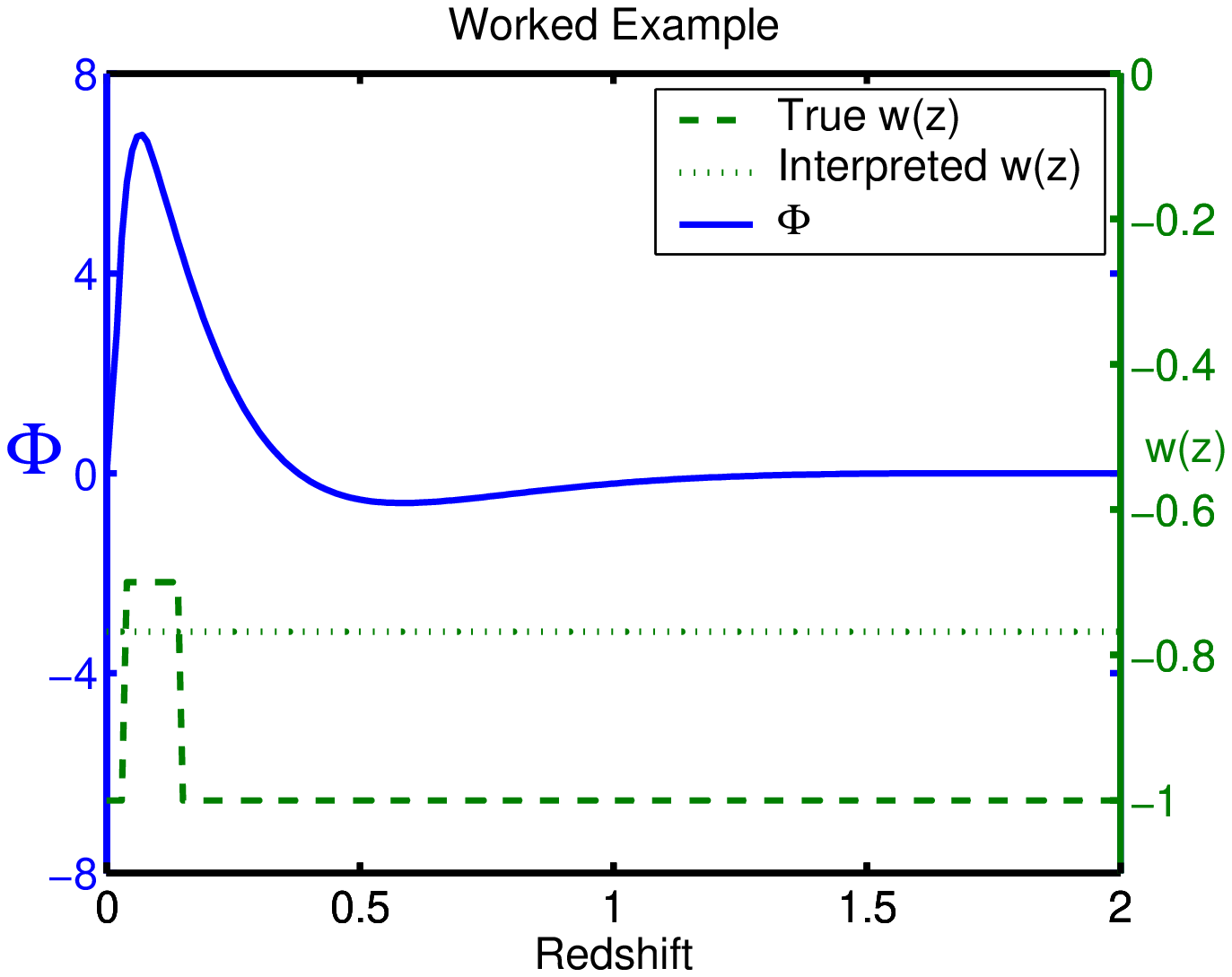}
\hspace{1cm}
\includegraphics[width=3in]{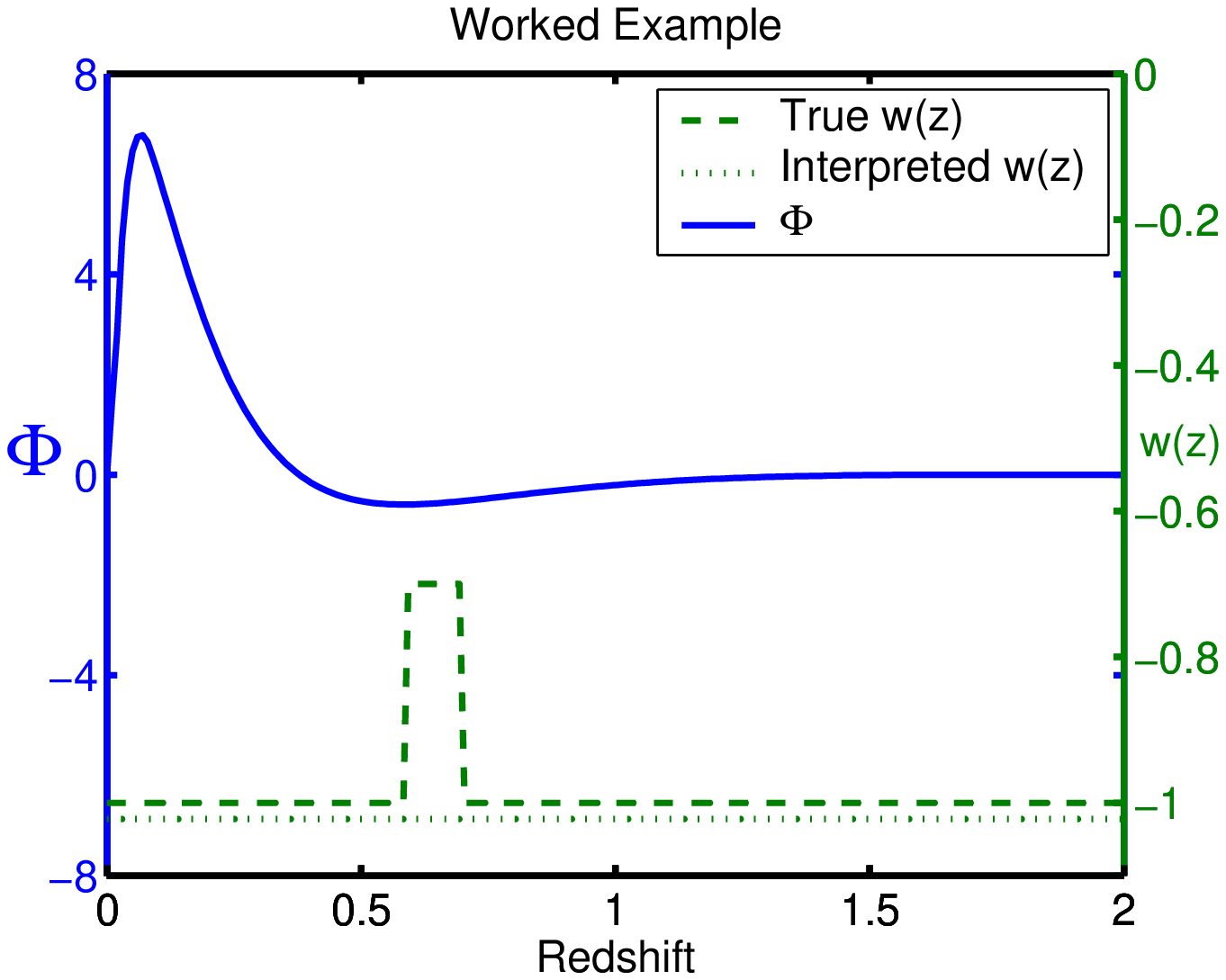}
\caption{ \emph{Left:} \label{fig:Work1} This is an example
illustrating the meaning of the weight function. A perturbation in
the underlying $w(z)$ model (dashed) leads to a shift in our best
fit (dotted) with a magnitude given by the corresponding value of
the weight function at that redshift. \emph{Right:}
\label{fig:Work2} A similar scenario to Fig.~\ref{fig:Work1},
except here the perturbation in the true $w(z)$ (dashed) occurs in
a region of negative weight. We therefore find that the best fit
value for $w$ (dotted) is pushed in the \emph{opposite}
direction.}
\end{figure*}

The worked examples in Fig.~\ref{fig:Work1} are designed to
highlight the meaning of the weight function, with two different
underlying $w(z)$ functions chosen for illustrative purposes. In
each case the dark energy model (dashed line) has $w(z)=-1$, with
a positive perturbation to $w=-0.7$ at some redshift. In the first
example this jump is at redshift $z \sim 0.1$ where the weight
function is large and positive. In this case fitting $w(z)=w_0$
gives a $w_0$ which is larger than $-1$, as may be qualitatively
expected by averaging the underlying $w(z)$ model. The extent of
the deviation in the best-fit value (dotted line) is proportional
to the height of the weight function.

What does it mean for a weight function to have negative regions?
In the second worked example the perturbation is at redshift $z
\sim 0.6$ where the supernova weight function is negative. In this
case the fitted $w_0$ is \emph{less} than $-1$, which seems
counterintuitive given that $w(z)$ is greater than or equal to
$-1$ at all times. However, we can still see that the fitted $w_0$
is a weighted average of the underlying $w(z)$, as described by
the weight function $\Phi(z)$.

As mentioned earlier, this effect was already demonstrated by Maor
et al, who used $w(z)=-0.7 + 0.8z$ and showed that this gives a
fit of $w \sim -1.75$, with the entire $95\%$ contour lying at
$w<-1$. Here we have provided a way of quantitatively predicting
this effect, provided the deviation from the fiducial model is not
too large. It arises from uncertainty in the value of both
$\Omega_{\rm m}$ and the ``nuisance" parameter $\mathcal{M}$ (we
discuss the effect of the priors later). For example,  a
$\Lambda$CDM cosmology with an enhanced value of $\Omega_{\rm m}$
reproduces an expansion history nearly identical to one in which
$w(z)$ increases with redshift.

\subsection{Cosmic Shear}

Dark energy modifies the geometry and strength of the lenses
contributing to the weak lensing seen in cosmic shear. Here we
consider a high redshift survey with source galaxies divided into
two redshift bins. Survey parameters correspond to those used by
Refregier et al. \cite{2004AJ....127.3102R} for a SNAP-like
mission. Note that our results are insensitive to the area of sky
covered. Here we marginalise over $\Omega_{\rm m}$, $\Omega_b$,
$n$, $h$ and $\sigma_8$, each with a flat prior.

In contrast to supernovae, the first two eigenvalues are both
quite significant for cosmic shear. The weight function is mostly
positive, with a single broad peak at a redshift around 0.5, and a
mean readshift of 0.58. The redshifts probed by cosmic shear are
higher than those for a similar redshift supernova survey; in a
previous paper \cite{2005PhRvD..71h3501S} we attributed this to
the combination of effects on both the geometry of the Universe
and the growth of structure.

\subsection{Baryon Oscillations}

The distance travelled by sound waves prior to recombination, $s$,
is a standard ruler embedded within both the matter power spectrum
and the anisotropies of the cosmic microwave background (CMB).
Galaxy surveys will be able to measure this distance as it appears
on the sky, probing the angular diameter distance to the survey
redshift. Provided spectroscopic redshifts are available, its
appearance along the line of sight can be used to determine the
Hubble parameter at the redshift of the survey to within a few
percent. Thus, like supernovae, this is a purely geometrical test
of dark energy. For a recent discussion of this approach, and the
potential costraints on dark energy, see Glazebrook \& Blake
\cite{2005ApJ...631....1G}.

Whilst the survey parameters of WFMOS/KAOS
\footnote{http://www.noao.edu/kaos/} are yet to be confirmed, a
guideline survey is adopted, covering $1000$ square degrees at low
redshift ($0.5<z<1.3$) and $400$ at high redshift ($2.5<z<3.5$).
We consider the observables for each redshift bin to be the values
of $D_A(z)/s$ and $H(z)/s$, averaged over the bin. Baryonic and
dark matter densities determine $s$, as outlined in Eisenstein \&
Hu \cite{1998ApJ...496..605E}, so we marginalise over these, along
with the Hubble parameter $h$. A Planck-like prior is included
whereby $D_A(z=1100)/s$ is determined with a precision of $0.2\%$,
and $\Omega_bh^2$ to $2\%$. We also find it necessary to include a
prior of 0.05 on $\Omega_{\rm m}$, as it is unlikely any
competitive constraints on $w$ could be produced without this
extra information. In any case, the consequences of relaxing and
strengthening this prior can be found in the following section.

We utilise the scaling relation from Blake et al.
\cite{2006MNRAS.365..255B} to evaluate the errors in $D_A(z)/s$
and $H(z)/s$. The redshift dependence of these errors exerts a
significant influence on the form of $\Phi(z)$. As with
supernovae, the PCA eigenvalues for baryon oscillations fall off
quite sharply, with the result that the weight function closely
resembles the first eigenmode. It peaks at zero, but maintains a
high level of sensitivity due to the source galaxies extending out
to $z=3.5$, and information from the CMB. This results in a mean
redshift of $0.54$.

Why is the form of all the weight functions so different? This
mainly arises from unique features in each, such as the ``nuisance
parameter" for supernovae, while the baryon oscillations have a
direct handle on the Hubble parameter at early times. The
cosmological dependence of the sound horizon $s$ is also a
significant factor. Cosmic shear on the other hand, is sensitive
to the growth of density perturbations in addition to the
geometric effects.

\subsection{Combined}

Now we assess the consequences of a combined dataset, whereby the
survey parameters are modified such that each can independently
determine $w$ to the same level of accuracy ($10\%$). This can be
seen as the dot-dash line in the bottom right hand panel of
Fig.~\ref{fig:Const}. In this case, the dominant feature is
attributed to the decay of dark energy toward higher redshift. It
should be noted that throughout this work the absolute errors
involved are unimportant, rather it is the relative strengths
which determine the form of the weight function.

\subsection{Supernova Legacy Survey}

\begin{figure}
\includegraphics[width=3in]{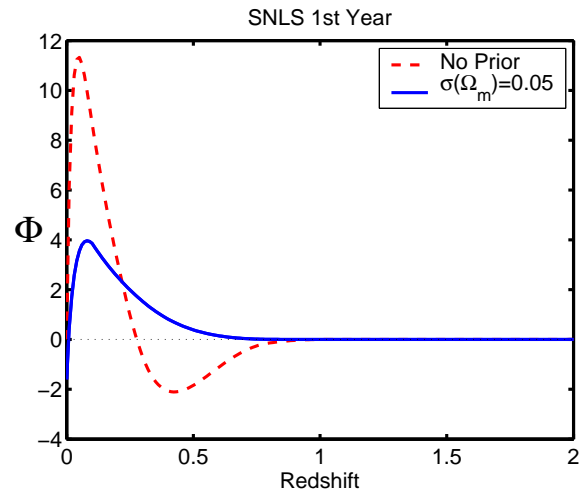}
\caption{\label{fig:snls}The weight function for the first year
results of the Supernova Legacy Survey.}
\end{figure}

\begin{figure*}
\includegraphics[width=5.8cm]{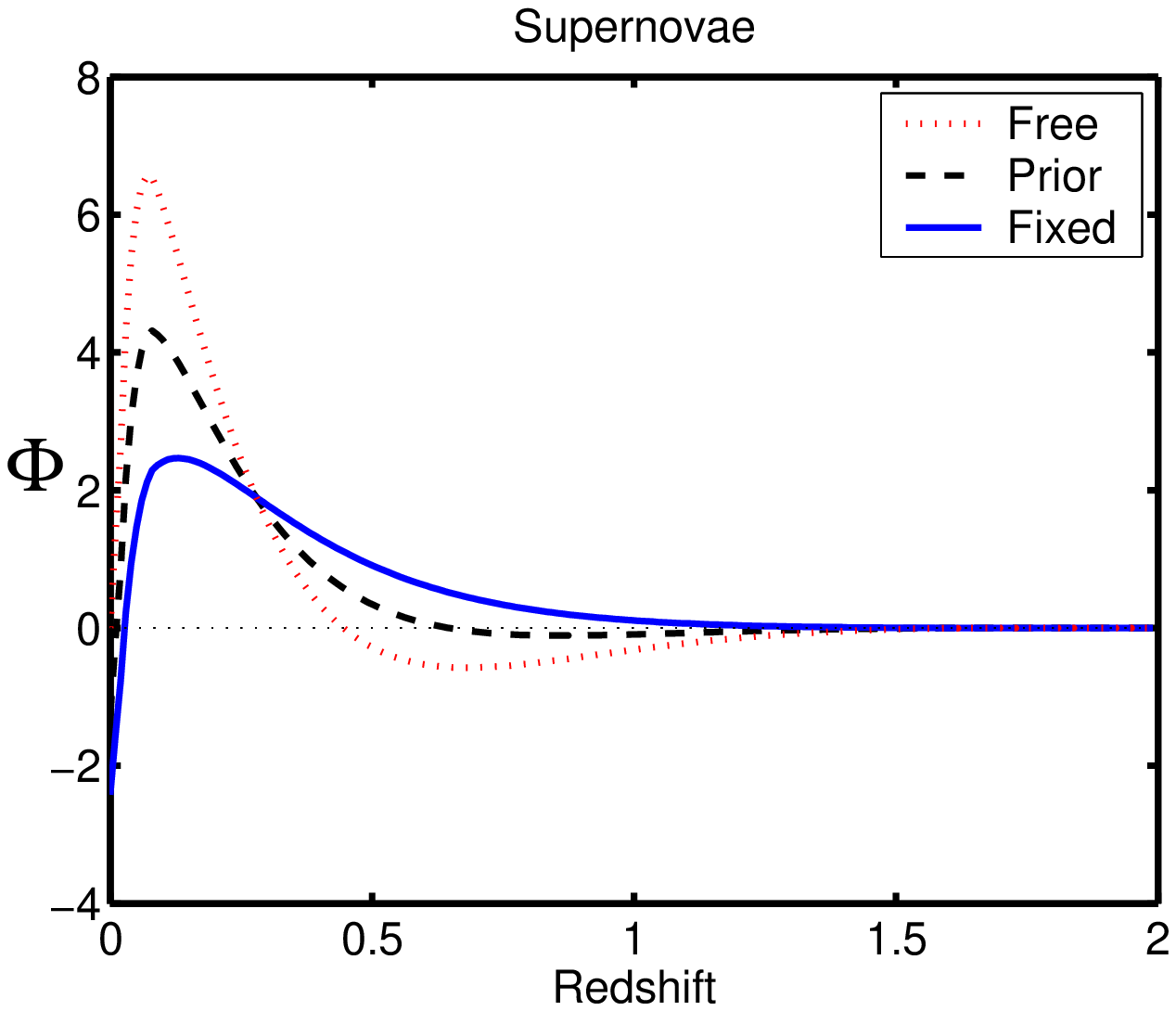}
\includegraphics[width=5.8cm]{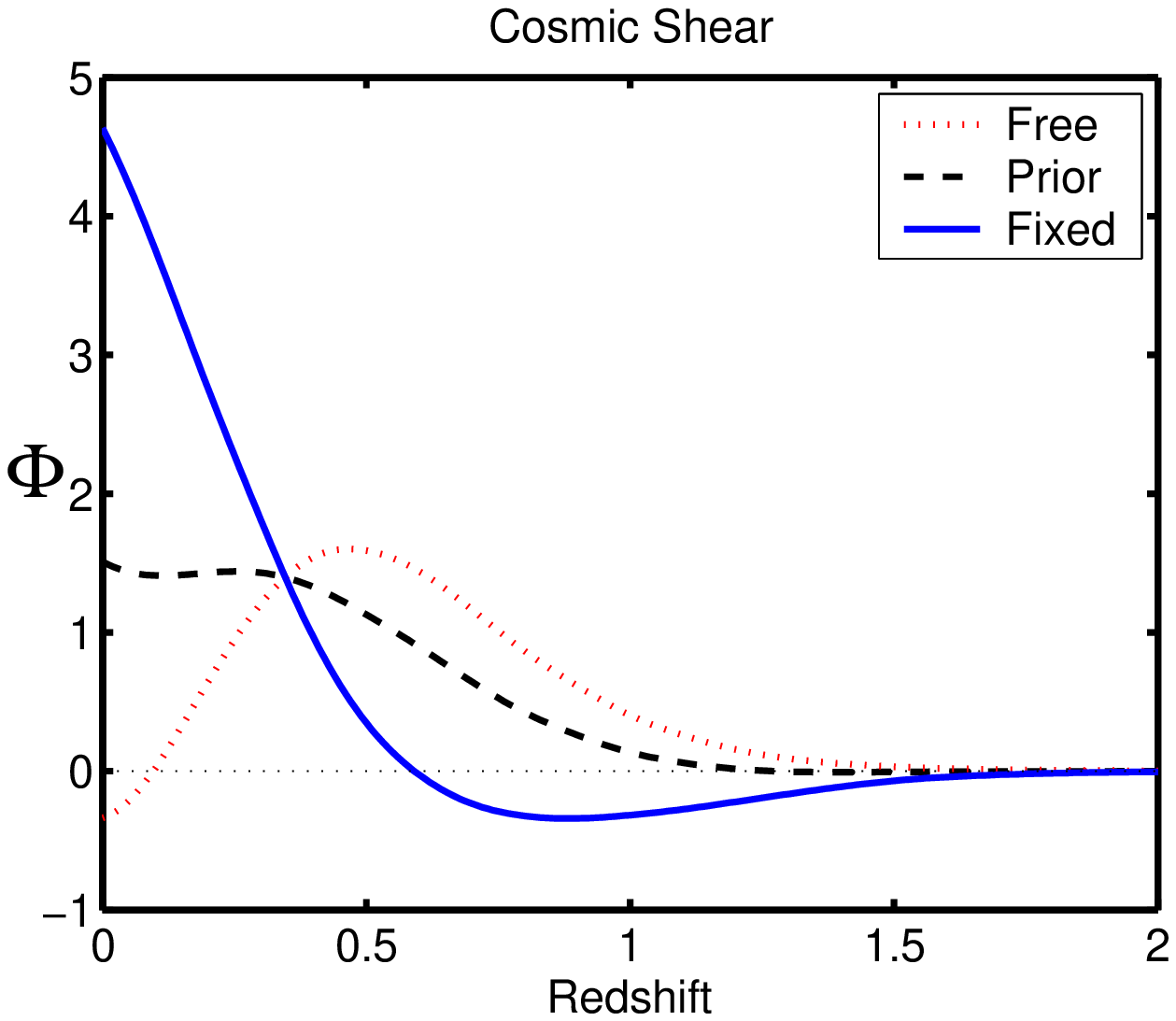}
\includegraphics[width=5.8cm]{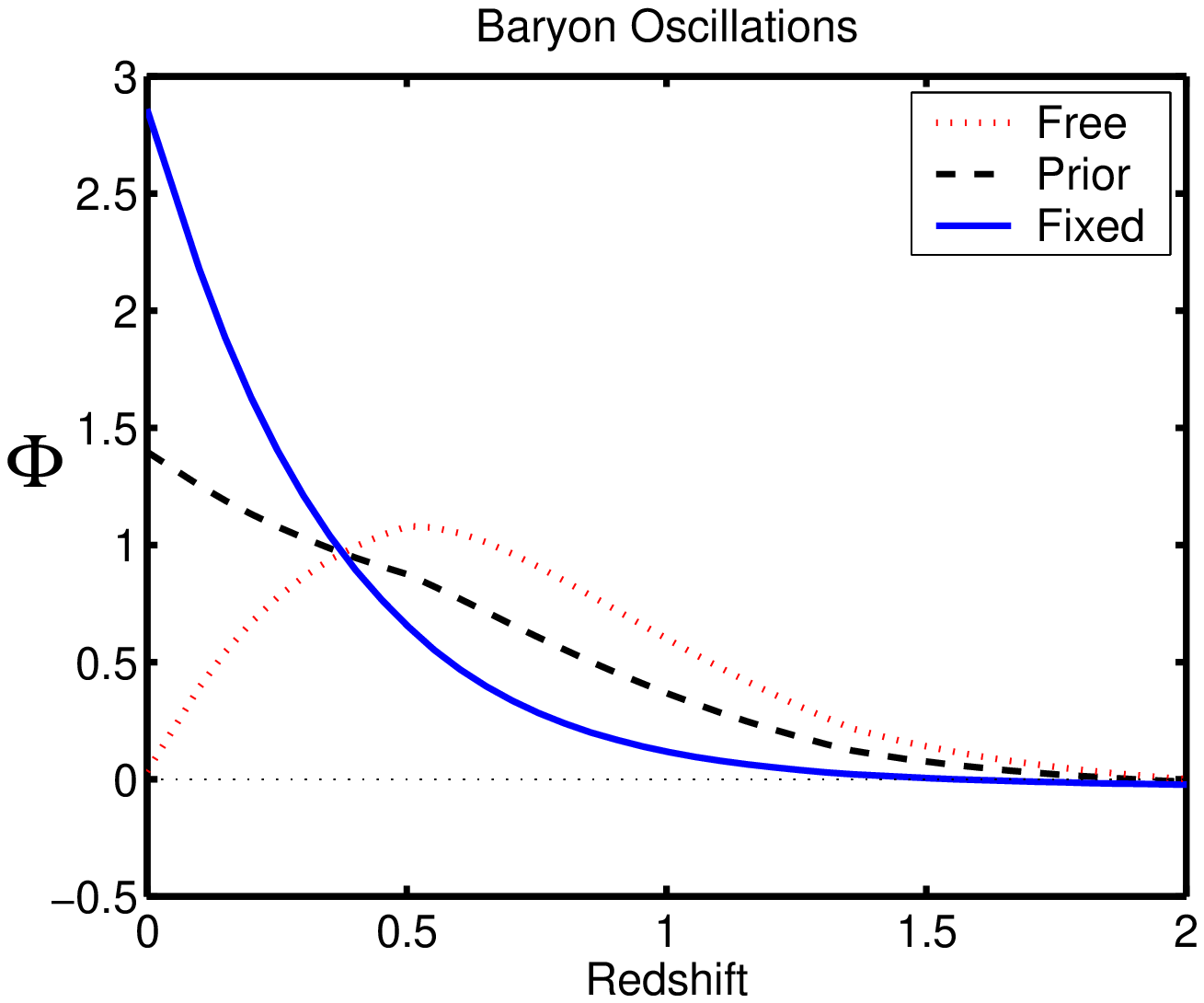}
\caption{ \label{fig:OmegaMsnae} Here we see the influence of the
$\Omega_{\rm m}$ prior. \emph{Left:} Supernovae lose the region of
negative weight at high redshift if $\Omega_{\rm m}$ is known
correctly (solid), although it does recur at very low redshift.
\emph{Center:} \label{fig:OmegaMshear} For cosmic shear there is a
significant change in redshift sensitivity when adding a stronger
prior on $\Omega_{\rm m}$ (dashed and solid). Contributing
information at redshift zero strengthens the influence of $w$ at
low redshifts. Thus fixing the present day matter density raises
the sensitivity at later times. \emph{Right:}
\label{fig:OmegaMbao} The weight function for baryon acoustic
oscillations change in a similar way to cosmic shear when placing
the weakest prior on $\Omega_{\rm m}$ (dotted).}
\end{figure*}

We have been considering the weight functions of the potential
surveys of the future, as a means to compare the techniques.
However, it is also of interest to assess the meaning of
present-day constraints. The weight function corresponding to the
results from the first year of the Supernova Legacy Survey (SNLS)
\cite{astier-2005-} is shown in Fig.~\ref{fig:snls}.

This data alone is insufficient to break the degeneracy between
$w$ and $\Omega_{\rm m}$, as shown in Fig. 6 of
\cite{astier-2005-}. On adding a Gaussian prior on $\Omega_{\rm
m}$ of width 0.05 centered on the true value we establish a mean
redshift sensitivity of 0.19.

From Fig. 6 of \cite{astier-2005-} we can see that applying a
prior of $\Omega_{\rm m} =0.25 \pm 0.05$ would give a similar $w$
constraint to that given in their abstract from combining SNLS
with baryon oscillations. Thus 
with this prior on $\Omega_{\rm m}$ we can \emph{already} say that
the average value of $w(z)$ between $0<z \lesssim 0.4$ is
$\sim-1$, irrespective of its actual functional form. Conversely,
we have learnt nothing of the value of $w(z)$ beyond $z \gtrsim
0.4$, despite the redshift distribution of supernovae extending
out to $z=1$.

\section{Influence of Priors}
\label{sec:Prior}

In the above we have, for the most part, marginalised over the
cosmological and nuisance parameters with a flat prior on each,
therefore assuming minimal knowledge of their values. However the
shapes of the PCA eigenmodes and weight functions are expected to
depend on the exact priors applied. In this section we demonstrate
the variability of the weight functions by re-calculating them for
different priors. Note that our previous paper
\cite{2005PhRvD..71h3501S} includes results for supernovae and
cosmic shear where the cosmological and nuisance parameters are
known exactly (delta function priors).

The tightest constraints on $w$ will arise from supplying
additional datasets such as surveys of large scale structure, and
the CMB. This will help break degeneracies with cosmological
parameters. The most significant shift in the shape of the weight
function is found to arise from applying a prior to $\Omega_{\rm
m}$, and therefore we focus our discussion on this. Different
priors on $\Omega_{\rm m}$ cause the weight function to shift
between the forms seen in Fig.~\ref{fig:OmegaMsnae}.

For each cosmological probe we compare

(i) ``Free" (dotted line): a flat prior on $\Omega_{\rm m}$ (and
flat priors on all other cosmological parameters)

(ii) ``Prior" (dashed line): a Gaussian prior on $\Omega_{\rm m}$,
centered on the true value, and of a width comparable to the
constraint on $\Omega_{\rm m}$ placed by that particular survey.
Specifically, $\sigma(\Omega_{\rm m}) = 0.03, 0.01, 0.05$ for the
supernova, cosmic shear and baryon acoustic oscillations
respectively. Note that if the survey areas were changed then the
relative power of these $\Omega_{\rm m}$ priors would also change.

(iii) ``Fixed" (solid line): a delta function prior on
$\Omega_{\rm m}$ at the true value.

In all cases the weight function using a Gaussian prior lies
mid-way between the two extremes. For cosmic shear and baryon
oscillations, the additional information on the matter density at
redshift zero naturally raises low-redshift sensitivity.
Supernovae break this pattern due to the uncertainty of their
absolute magnitude, which prevents us from studying the very
recent behaviour of dark energy. Indeed, without calibration
surveys such as the Nearby Supernova Factory
\cite{2002SPIE.4836...61A} the uncertainty on $\mathcal{M}$
increases, and this pushes the peak to slightly higher redshift.

\begin{figure*}
\includegraphics[width=5.5cm]{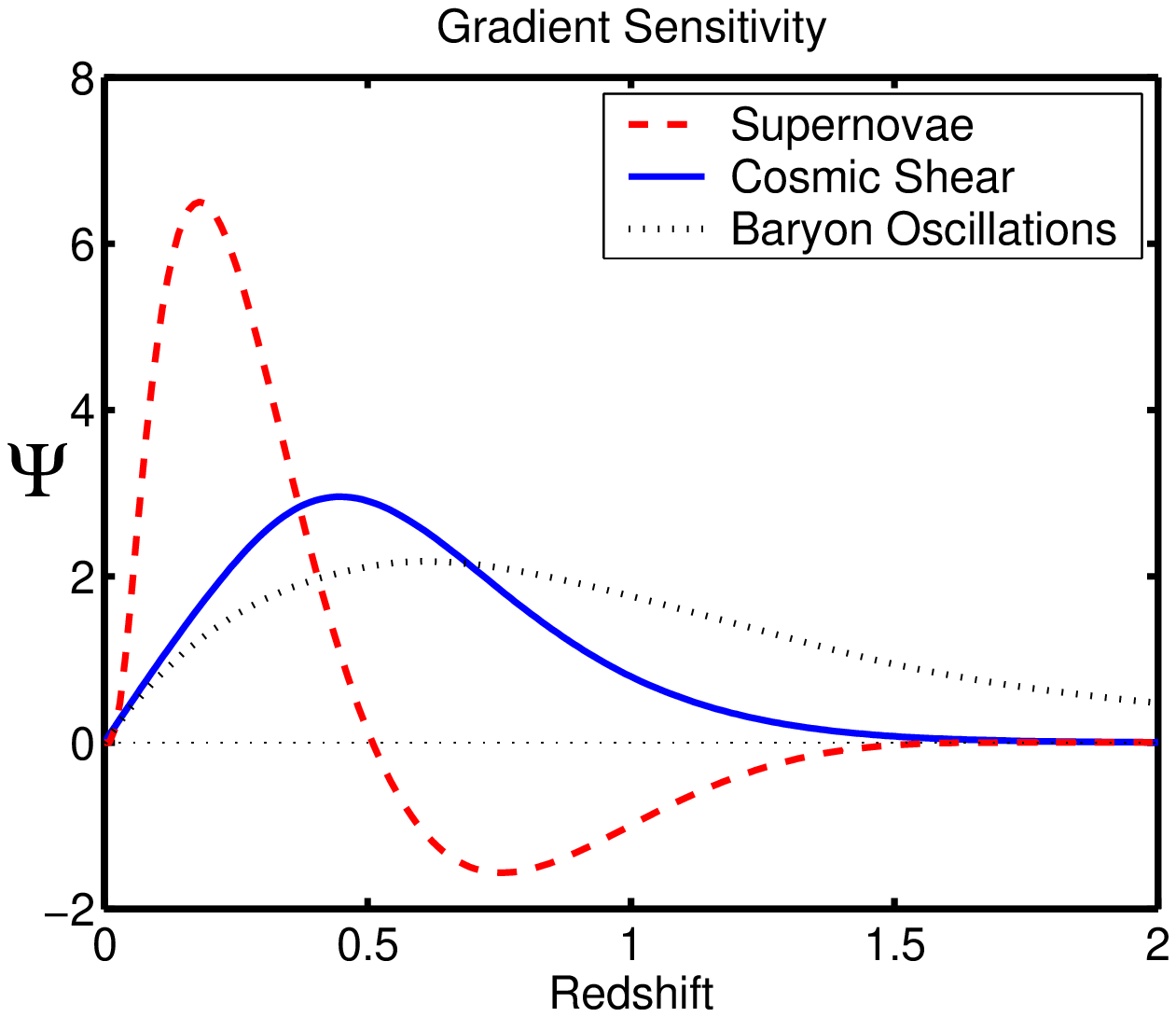}
\hspace{-0.2cm}
\includegraphics[width=5.8cm]{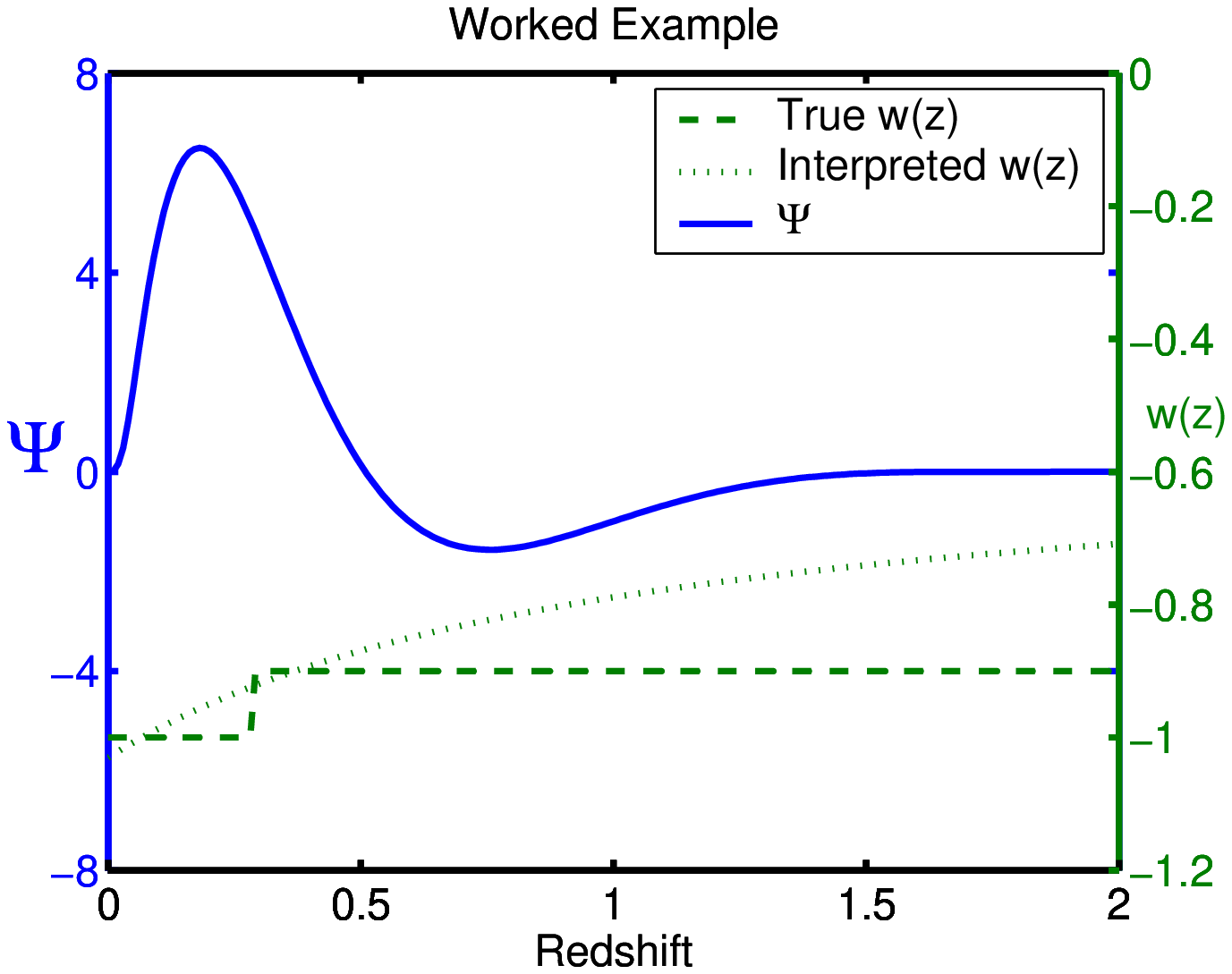}
\hspace{-0.2cm}
\includegraphics[width=5.8cm]{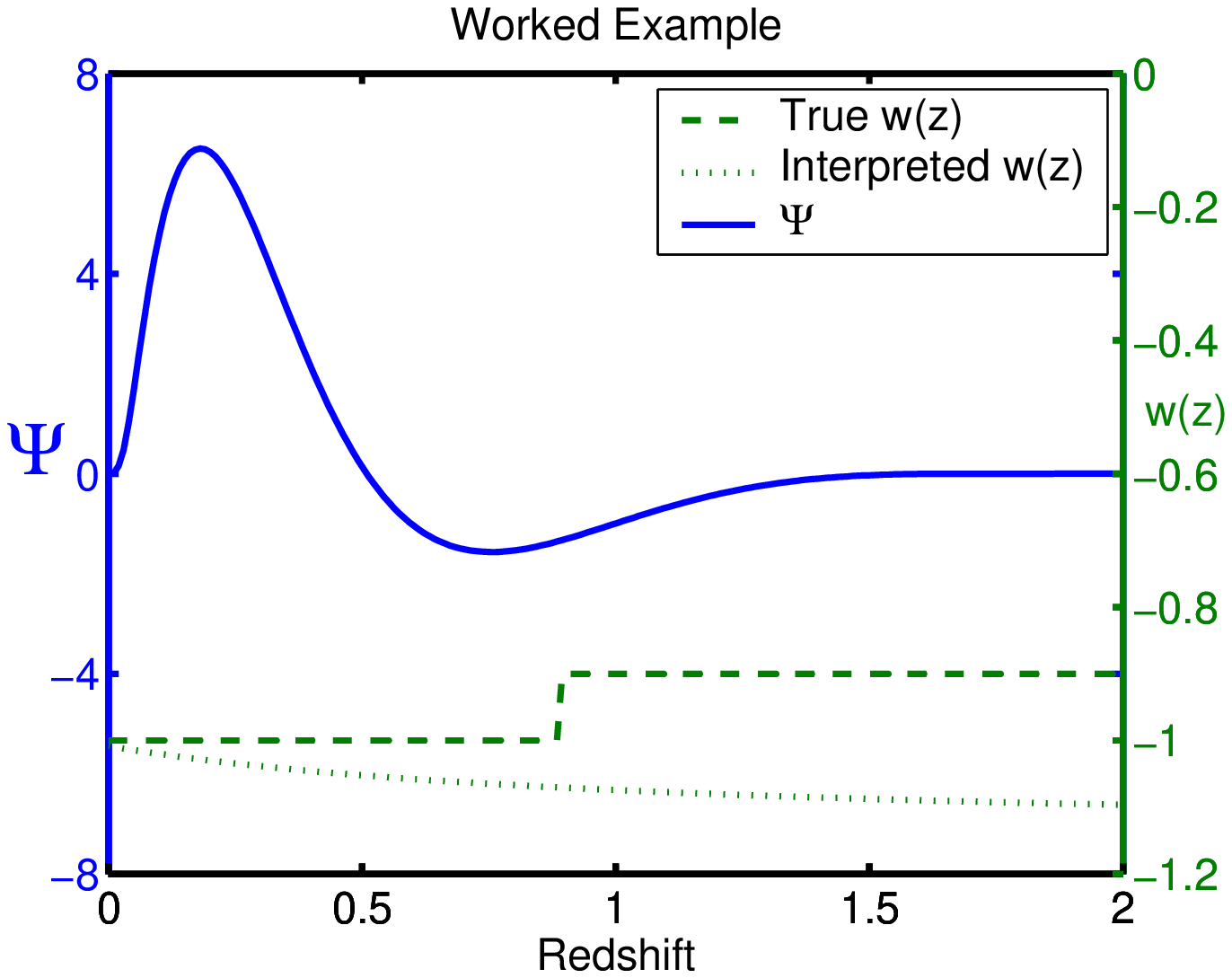}
\caption{ \emph{Left:} \label{fig:Grad} The gradient weight
function for each probe. These lines represent the sensitivity to
which each technique measures the first derivative of $w$, when
adopting the standard parameterisation $w=w_0+w_a(1-a)$.
\emph{Center:} \label{fig:WorkGrad1} This illustrates the
interpretation of the gradient weight function, $\Psi$. The solid
line shows the supernova gradient weight function; the dashed line
shows the true $w(z)$ used to simulate supernova magnitudes; the
dotted line shows the best fit $w(z) =w_0 + w_a (1+a)$ model to
the simulated data. \emph{Right:} \label{fig:WorkGrad2} The only
gradient in $w(z)$ occurs at $z \sim 1$, and since this occurs in
a region of negative weight, we find that the interpreted gradient
is significantly misleading.}
\end{figure*}

\section{Higher-order Parameterisation}
As data improves we are likely to advance from the
single-parameter model of a constant equation of state, and look
towards constraining some form of evolving equation of state. Here
we shall consider the standard parameterisation $w=w_0+w_a(1-a)$
(although the results are qualitatively unchanged using the
parameterisation $w=w_0 + w_1 z$).
Adopting an analogous approach to the previous section, we define
the \emph{gradient} weight function
\begin{equation} \label{eq:wgrad}
w_a^{fit}=\int \Psi(z) \frac{\ud w}{\ud z}(z) \ud z .
\end{equation}
\noindent The purpose of this new weight function is to quantify
how much we learn about the \emph{variation} of $w(z)$ from the
fitted parameter values. In this section we focus on fitting $w_a$
to the data. The left hand panel of Fig.~\ref{fig:Grad} represents
the redshift sensitivity with which we observe the gradient of $w$
for each of the three probes (see Appendix \ref{sec:scalefactor}
for an equivalent plot recast in terms of the sclae factor). The
gradient weight function is intended to be a fairly intuitive
tool. To illustrate the point of introducing $\Psi(z)$, the
sensitivity to change in the equation of state, two examples are
given in the centre and right hand panels of
Fig.~\ref{fig:WorkGrad1}. In these examples we select a true
equation of state in the form of a step function. The purpose of
choosing this $w(z)$ is to localise all the gradient in one place,
and see how the fitted value of $w_a$ responds. The value of the
$\Psi(z)$ at the location of the step will determine our best fit.

In the central panel of Fig ~\ref{fig:WorkGrad1} the true $w(z)$
changes from $w=-1$ at low redshift ($z<0.4$) to $w=-0.8$ at
higher redshift ($z>0.4$). Since the gradient of $w(z)$ is
everywhere zero, except for a positive delta function at $z=0.4$,
the resulting best-fit is readily determined from Equation
\ref{eq:wgrad}. It is simply the magnitude of the step, multiplied
by the value of $\Psi(z)$ at the location of the step.

The final plot within Fig ~\ref{fig:WorkGrad2} shows an example in
which the step occurs where the gradient weight function is
negative. As a result, the estimated gradient is a qualitatively
inaccurate representation of the underlying physics.

The gradient weight function provides a quick way of predicting
the fitted $w_a$ value for a given $w(z)$ model. Conversely, for a
given $w_a$ obtained by some experiment, we can now interpret it
using the gradient weight function. If $w_a$ is found to be
positive then either there exists (i) a positive gradient in the
true $w(z)$ in a region where the gradient weight function is
positive; or (ii) a negative gradient in the true $w(z)$ in a
region where the gradient weight function is negative; or some
combination of the two.

Since the supernova weight function does have a negative region,
this means we \emph{cannot} immediately interpret a positive $w_a$
to imply that $w(z)$ increases with redshift. At $z>0.5$,
perturbations in $\ud w / \ud z$ reverse the value for $w_a$.

Neither cosmic shear nor baryon oscillations possess significant
negative weight, allowing a more straightforward interpretation of
results. They are both most sensitive to changes in the true
equation of state at around a redshift of 0.5, although here it is
baryon oscillations which offer the highest redshift sensitivity
to transitions in $w$. Mean redshift sensitivities are 0.42,0.56,
1.14 and  for supernovae, cosmic shear and baryon oscillations
(again, using $|\Phi(z)|$ due to the negative weight). All
techniques lack sensitivity at redshift zero due to there being no
physical consequence to its present value.

On combining the three probes we obtain a combined gradient weight
function that resembles an average of the three separate gradient
weight functions (not shown here).

We consider the gradient weight function $\Psi(z)$ to provide a
more intuitive interpretation of $w_a$, although we could instead
write $w_a = \int \Phi_a(z) w(z) dz$ as for the standard weight
function (Eq.~\ref{eq:phi}). We show in Appendix
\ref{sec:Gradient} that two weight functions are trivially related
through

\begin{equation} \label{eq:phia}
\Psi(z)=-\int_0^z \Phi_a(z') \ud z' .
\end{equation}

\begin{figure}
\includegraphics[width=3in]{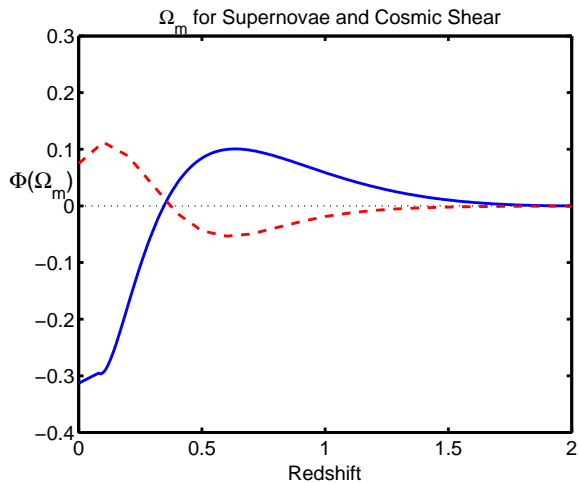}
\caption{\label{fig:OmegaM} The $\Omega_m$ weight function
quantifies the deviation in the fit of $\Omega_{\rm m}$ resulting
from redshift variations in $w$. This plot shows that if we fit a
constant $w$ when it actually increases with redshift, we will
obtain an overestimated value for $\Omega_{\rm m}$ with supernovae
(solid) but an underestimate with cosmic shear (dashed).}
\end{figure}

\section{Mistaken matter density}
So far, we discussed how the fitted values of a parameterised
equation of state relate to the true function. This is of
particular concern when the parametric form is a poor
representation of the true equation of state. A further
consequence may be the production of misleading values of the
other cosmological parameters. As the most important example, we
focus our attention on $\Omega_{\rm m}$.

In the work mentioned earlier, Maor et al found that fitting a
constant equation of state to data simulated from $[w(z)=-0.7 +0.8
z, \Omega_{\rm m}=0.3]$, the best fit values were $[w_0=-1.75,
\Omega_{\rm m} =0.65$]. Of course, this strongly erroneous value
of $\Omega_{\rm m}$ could be ruled out independently.  But when on
a less dramatic scale this effect will persist, and becomes more
difficult to exclude.

Here we quantify the apparent deviation in $\Omega_{\rm m}$
induced by fitting a constant equation of state to the data, when
the true equation of state is actually evolving.
\begin{equation} \label{eq:PhiOm}
\Omega_{\rm m}^{\rm fit}=\Omega_{\rm m}^{\rm true}+\int
\Phi_{\Omega_{\rm m}}(z) \delta w(z) \ud z .
\end{equation}
\noindent This gives the fitted value of $\Omega_{\rm m}$ in terms
of the true value, and the deviation of the equation of state
$\delta w(z)$ from the constant $w$ model.

The weight function $\Phi_{\Omega_{\rm m}}(z)$ is defined by this
equation, but is only valid for small perturbations about the
fiducial model. Other parameter's fits may also be skewed by
perturbations in $w(z)$, and the general expression for the weight
function of a given parameter is outlined in Appendix
\ref{sec:General}. Note that these weight functions integrate to
zero, as expected since the parameter is correctly estimated if
the true equation of state is a constant.

Fig.~\ref{fig:OmegaM} compares the sensitivity with which a
perturbation in $w$ will be interpreted as a change in the
best-fit value of $\Omega_{\rm m}$. The responses of supernovae
and cosmic shear are similar, but of opposite sign. Results from
baryon oscillations are omitted as they are unable to provide
competitive constraints  on $\Omega_{\rm m}$, for the survey
parameters considered here.

The supernova $\Omega_{\rm m}$ weight function is initially
negative, becoming positive beyond $z \sim 0.4$. Consider the
outcome of Eq.~\ref{eq:PhiOm}, for the case of an equation of
state with positive gradient. The fitted $\Omega_{\rm m}$ value
will be \emph{larger} than the true value of $\Omega_{\rm m}$.
This qualitatively fits the result of Maor et al. The quantitative
deviation is inaccurate since such large deviations from the
fiducial model leads to a major underestimation of the weight
function's amplitude at high redshift.

The cosmic shear $\Omega_{\rm m}$ weight function has the opposite
shape, a good illustration of the complementarity with supernovae.
So a positive gradient in $w(z)$ leads to an \emph{underestimated}
$\Omega_{\rm m}$ if fitting a $w={\rm const}$ model. The amplitude
of the cosmic shear weight function is smaller, so this effect is
slightly weaker, due to its ability to constrain $\Omega_{\rm m}$.

\section{Discussion}

For the foreseeable future, a full reconstruction of $w(z)$ is out
of reach, but we can significantly narrow the family of functions
which remain compatible with observation. We have seen how
constraints from three dark energy probes, each capable of placing
percent-level constraints on $w$, respond to a general $w(z)$.

The weight functions produced offer a compact and intuitive way of
characterising fitted parameters such as $w_0$ and $w_a$. For a
SNAP-like supernova survey sensitivity to dark energy typically
peaks at $z \sim 0.05$, but this peak may be at slightly higher
redshift either when applying a prior on $\Omega_{\rm m}$, or when
lacking a local sample. Mean redshift sensitivities are 0.28,
0.58, and 0.54, for supernovae (SNAP-like), cosmic shear
(SNAP-like), and baryon oscillations (WFMOS).

Results from baryon oscillations offer the most straightforward
interpretation since their weight function is everywhere positive.
This is due to a combination of factors, including the direct
measure of the Hubble parameter, and the cosmological dependence
of the sound horizon. At high redshift, the sensitivity is
inevitably suppressed by the lack of dark energy. Indeed, at
redshifts beyond those considered here, it becomes more
appropriate to discuss constraints in terms of $\Omega_\Lambda$ as
opposed to $w$.

There is a further benefit to an everywhere-positive weight
function, besides a more intuitive interpretation of results. Let
us consider the constraints from SNLS, where $w=-1.023 \pm 0.09$.
Upon adding a theoretical prior that $w(z)$ does not traverse the
$-1$ boundary, the evolution of $w(z)$ becomes strongly confined,
since we know the average value must be close to $-1$. This would
not be the case if there were significant regions of negative
weight, since higher-order terms in the Taylor expansion of $w(z)$
would enable significant evolution away from minus one, whilst
maintaining a constant fit close to minus one. An alternative
perspective is that the family of dark energy models constrained
by a survey weight is more widely dispersed when the corresponding
weight function exhibits negative regions.

The choice of dark energy probe is an eagerly anticipated one. Of
primary concern is the level of reliability and accuracy of
constraints which could be attained. In the case of comparable
performance, one could then prioritise by considering the weight
functions to indicate those which probe preferential regions.
However, none of the approaches could be considered sufficiently
foolproof to act alone. For example, it is unlikely that any
detected deviation from $w=-1$ would become widely accepted until
there is independent verification.

\begin{acknowledgments}
We thank Steve Rawlings and Roger Blandford for helpful comments.
FRGS acknowledges the support of Trinity College. SLB thanks the
Royal Society for support.
\end{acknowledgments}

\begin{appendix}
\section{The Scale Factor}
\label{sec:scalefactor}
 Expressing our results in terms of the scale
factor $a=(1+z)^{-1}$ arguably provides greater clarity, and a
more physical representation of the expansion history. We define
the weight function in terms of the scale factor such that

\begin{equation} \label{eq:wgrada}
w_a^{fit}=\int \Phi(a) w(a) \ud a .
\end{equation}

\begin{figure} \includegraphics[width=3in]{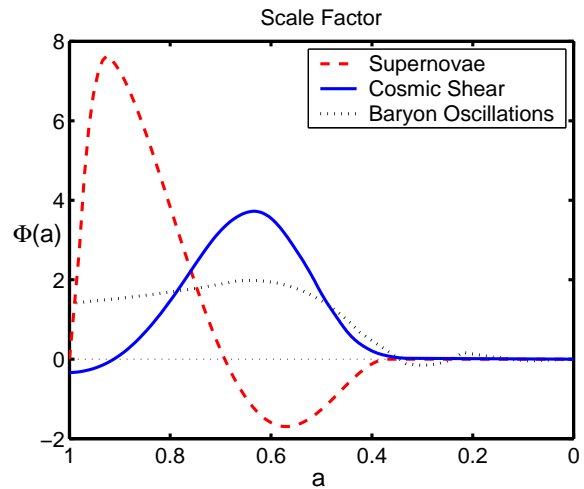}
\caption{\label{fig:scalefactor} Here we express the sensitivity,
when fitting a constant $w$, as a weighted average over the scale
factor.}  \end{figure}

This function is obtained by noting the relation $\Phi(a) \ud
a=\Phi(z) \ud z$. The plots shown in Fig.~\ref{fig:scalefactor}
are equivalent to those in Fig.~\ref{fig:Const}.  Note the galaxy
redshift bins for the baryon oscillations are now visible in the
form of the weight function.

\begin{figure} \includegraphics[width=3in]{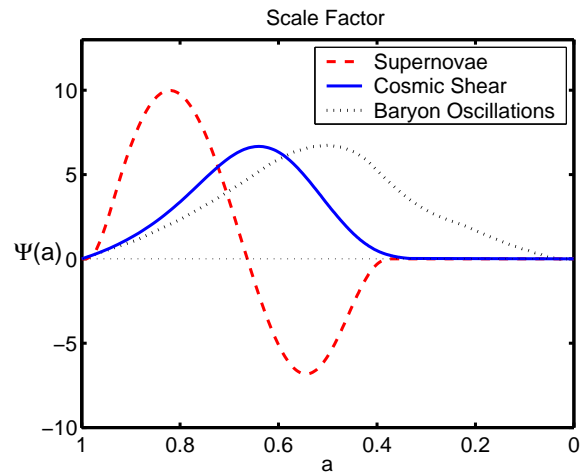}
\caption{\label{fig:scalefactorgrad} The sensitivity to the
gradient in $w$, as a weighted average over the scale factor.}
\end{figure}

In an extension of this approach, Fig.~\ref{fig:scalefactorgrad}
reproduces the gradient sensitivities to $w$, and is equivalent to
the left panel of Fig.~\ref{fig:Grad}.

\section{Cosmological Dependence}
\label{sec:w08}

\begin{figure}
\includegraphics[width=3in]{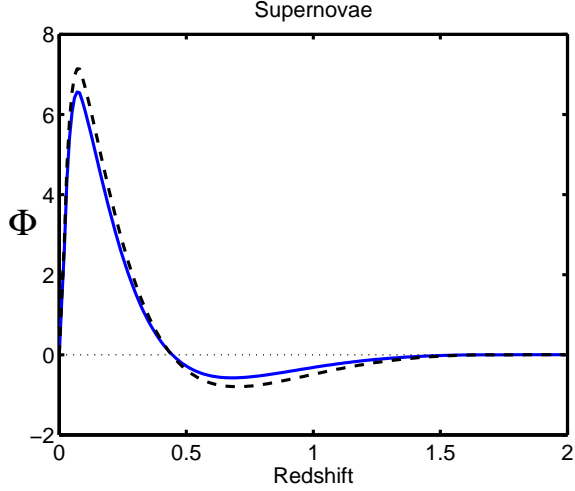}
\caption{\label{fig:w08} Here we see the enhancement of the weight
function produced when the fiducial model is $w=-0.8$ (dashed
line). This is pronounced at high redshift, where dark energy is
now more prevalent.}
\end{figure}

 Weight functions, as with PCAs, are
limited by their dependence on cosmological parameters. In
Fig.\ref{fig:w08}, the solid line is for a fiducial model with
$w=-1$ and the dashed line for a fiducial model with $w=-0.8$. We
see that a change in the fiducial value of $w$ preserves the form
of $\Psi$, but there is a moderate change in amplitude. If
investigating a particular model, it is likely this could be
compensated for by including a functional dependence of $w(z)$ on
the weight function, thereby scaling the amplitude in accordance
with the amount on dark energy present. For example, a scenario in
which $w$ progresses towards zero (as may be anticipated to avoid
fine-tuning issues) would lead to weight functions with amplified
power at high redshift.

\section{\label{sec:WtPCA} Weight Functions from PCAs}

Here we derive Eq.~\ref{eq:phipca} which relates the weight
function for a constant $w$ fit to the principal components. The
coefficients of the principal components defined in
Eq.~\ref{eq:pcas} provide a convenient representation of our
observations:

\begin{equation}
\chi^2 = \sum
\Big(\frac{\alpha_i^{true}-\alpha_i^{fit}}{\sigma(\alpha_i)}
\Big)^2
\end{equation}

\noindent and we use the orthonormality of the eigenmodes to
deduce

\begin{equation}
\alpha^{true}_i = \int w(z)  e_i(z) \ud z \,\,\,,
\end{equation}

\begin{equation}
\alpha^{fit}_i = w^{fit} \int e_i(z) \ud z \,\,\,.
\end{equation}

By minimising $\chi^2$, and solving for $w^{fit}$, the expression
for $\Phi(z)$ follows from our definition of the weight function,

\begin{equation} \label{eq:phiappendix}
w^{fit}=\int \Phi(z) w(z) \ud z \,\,\, ,
\end{equation}

\begin{equation}
\Phi(z)= \frac{\sum_i{e_i(z)\int e_i(z') \ud
z'/\sigma^2(\alpha_i)}}{\sum_j (\int e_j(z'') \ud z'')^2
/\sigma^2(\alpha_j)} \,\,.
\end{equation}

\noindent In practice, we divide $z$ into 200 bins, and therefore
the equivalent expression in matrix form becomes
\begin{equation}
\Phi(z_j)=\frac{\sum_{i,k}K_{ij}K_{ik}}{\sum_{i,k}K_{ik}K_{ik}}
\end{equation}

\noindent where we have defined

\begin{equation}
 K_{ij}=e_i(z_j)/\sigma(\alpha_i)
\end{equation}
(no summation).

\section{\label{sec:Gradient} The Gradient Weight Function}
We turn our attention to the derivation of the gradient weight
function, which quantifies the redshift at which the derivative of
$w(z)$ is measured,

\begin{equation} \label{eq:w1}
w_a^{fit}=\int \Psi(z) \frac{\ud w(z)}{\ud z} \ud z .
\end{equation}
The most rapid evaluation of $\Psi(z)$ again involves the PCAs,
and we only need consider the first few eigenmodes. We start by
defining the matrix
\begin{equation}
 K_{ij}=e_i(z_j)/\sigma(\alpha_i)
\end{equation}
\noindent as this concisely represents the data in terms of
observations with uncorrelated errors. When simultaneously fitting
two parameters, as with $w=w_0+w_a(1-a)$, we are left with a pair
of simultaneous equations, arising from the minimisation of
$\chi^2$. Therefore
\begin{equation}
\Phi_a=\frac{(uK^TKu^T)uL^TK-(uK^TLu^T)uK^TK}{(uK^TKu^T)(uL^TLu^T)-(uK^TLu^T)^2}
\end{equation}
\noindent where we have defined

 \begin{equation}
 L_{ij}=K_{ij}(1-1/(1+z_j))
\end{equation}
(no summation). This is essentially an extension of Equation 19
from Saini et al. \cite{2003MNRAS.343..533D}. We then convert to
the gradient form $\Psi(z)$ via (\ref{eq:phia}).

\section{\label{sec:General} The Generalised Weight Function}

Previously we have considered the PCAs corresponding to measuring
$w(z)$, when marginalising over other parameters. To learn of the
effects of fitting other parameters, we introduce them as an
extension of the principal components. Our eigenvalues are
redefined as

\begin{equation}
\alpha^{true}_i = \sum_j e_{ij} P^{true}_j \,\,\,,
\end{equation}

\begin{equation}
\alpha^{fit}_i = \sum_j e_{ij} P^{fit}_j \,\,\,,
\end{equation}

\noindent where the components of $P$ include our original $w$
bins as before, plus the extra parameters of interest. These
correspond to the $j$th component of the eigenmodes $e_{ij}$.

Thus repeating the procedure of the previous section, with
$\chi^2$ minimisation, we arrive at a set of $n$ simultaneous
equations, where $n$ is the number of fitted parameters. This is
most readily solved with the division of matrices.

\end{appendix}
\bibliography{bibs}
\end{document}